\documentclass[jsss, final]{copernicus}

\usepackage{lineno}

\begin{document}
\nolinenumbers


\title{Quantitative measurement of combustion gases in harsh environments using NDIR spectroscopy}

\author[a]{Christian Niklas}
\author[a,b]{Stephan Bauke}
\author[a]{Fabian Müller}
\author[a,c,d]{Kai Golibrzuch}
\author[a]{Hainer Wackerbarth}
\author[a]{Georgios Ctistis}
\affil[a]{Laser-Laboratorium Göttingen e.V., Hans-Adolf-Krebs-Weg 1, 37077 Göttingen, Germany}
\affil[b]{IAV GmbH, Entwicklungszentrum
Nordhoffstraße 5,
38518 Gifhorn
Germany}
\affil[c]{Max-Planck-Institute for Biophysical Chemistry, Am Fassberg 11, 37077 Göttingen, Germany}
\affil[d]{Institute for Physical Chemistry, Georg-August-University Göttingen, Tammannstrasse 6, 37077 Göttingen, Germany}


\runningtitle{Non-dispersive IR-spectroscopy in harsh environments}

\runningauthor{Christian Niklas et. al.}

\correspondence{Georgios Ctistis (georgios.ctistis@llg-ev.de)}

\received{}
\pubdiscuss{} 
\revised{}
\accepted{}
\published{}


\firstpage{1}

\maketitle

\begin{abstract}
The global climate change calls for a more environmental friendly use of energy and has led to stricter limits and regulations for the emissions of various greenhouse gases.
Consequently, there is nowadays an increasing need for the detection of exhaust and natural gases.
This need leads to an ever-growing market for gas sensors, which, at the moment, is dominated by chemical sensors.
Yet, the increasing demands to also measure under harsh environmental conditions pave the way for non-invasive measurements and thus to optical detection techniques.
Here, we present the development of a non-dispersive infrared absorption spectroscopy (NDIR) method for application to optical detection systems operating under harsh environments.
\end{abstract}

\copyrightstatement{TEXT}

\introduction 
In today's world, climate change is one of the most demanding problems for our modern society with significant technological challenges in various areas. One of the main species contributing to global warming, carbon dioxide (CO$_2$), has increased from a level of 310\,ppm in the year 1972 to 410\,ppm today \citep{CO2level}. Therefore, the control and limitation of CO$_2$ is gaining importance, resulting in the need for gas detectors that are able to measure corresponding concentrations with sufficient precision at the location of emissions. 
Furthermore, besides environmental control, safety as well as process and quality control are also important applications for gas sensors. 
A well known safety issue is the control of combustion gases of civil fireplaces. Here, carbon monoxide (CO) is one of the most relevant gas species besides CO$_2$. The odorless CO is extremely toxic due to its chemical property to bind strongly to hemoglobin and therefore inhibiting oxygen transport \citep{ernst1998carbon}.  
Rooms with concentrations above 30 ppm are deemed hazardous for a person's health for a longer exposure \citep{BAuA2018}.

Another important area for CO$_2$ detection is exhaust emission control in the automobile and transport sector. Here, forthcoming new limits for CO$_2$ and NO$_x$ emission set by the European Union \citep{EU-Verordnung} force the development of more efficient and cleaner engines. Both are usually achieved by controlling and optimizing the combustion process, i.e. mixture formation prior to ignition (\cite{Grosch_2014}, \cite{Bauke2018}, \cite{Golibrzuch2017}).
A typical approach to reduce NO$_x$ emissions is the use of exhaust gas recirculation to lower combustion temperature. EGR-rates can be determined by monitoring CO$_2$ concentrations \citep{Grosch_2014}.

In contrast to sensors for civil applications requiring low-cost solutions, systems for engine development application are less price-sensitive but have, on the other hand, more demanding requirements. 
For example, they need to offer µs time resolution to enable crank-angle resolution and resolve single engine cycles (\cite{Grosch_2014}, \cite{Bauke2018}).

To date electrochemical and resistive sensors dominate the market for gas sensors \citep{trends2014trends}. Electrochemical sensors use two or three electrodes and reduce or oxidize the target gas and measure the resulting electrical current allowing a cheap detection method \citep{stetter2003sensors}. Nonetheless, these sensor types face various problems, e.g. limited durability due to the electrolyte or susceptibility for different gases \citep{electrochemical}. Here, hydrogen sulfides can influence the measurement of CO$_2$, which is especially dangerous for sewer measurements.
Furthermore, electrochemical sensors cannot be used in the environment of an internal combustion (IC) engine, as it is prone to the harsh environment and not capable of a high time resolution needed to analyze the mixture process of the fuel.

Where electrochemical sensors face usage limitations, application of optical sensors is often advantageous.
In combustion diagnostics, a frequent approach is the use of laser-induced fluorescence (LIF) for measurements of temperature or fuel-concentrations with high spatial resolution.
However, LIF measurements require sufficient optical access, do not allow real-time resolution, and measurements are time consuming due to the complex experimental set-up (\cite{Schulz_2005}; \cite{Luong_2008}). Most important, most gas-phase molecules cannot be excited to appropriate electronic states, so measurements rely on the use of fluorescent markers (tracer) that represent the species of interest.

Instead, non-dispersive infrared (NDIR) spectroscopy can be utilized for both civil fireplaces as well as IC engines, where gas specific infrared absorption spectra, present in almost any molecule, are used to determine the density of a gas.

In this work, we present the development of two sensors based on NDIR spectroscopy in harsh environments: (1) a low-cost sensor for civil fireplaces and (2) a high-speed sensor for determination EGR-rates in IC engines. Thereby, the sensors face following difficulties: a simultaneous measurement of two gases at different concentrations and simultaneous measurement of overlapping absorption of the analytes.

Prior to the presentation of details on the respective sensor systems, we introduce the basic principle of NDIR spectroscopy as well as the relevant spectroscopic properties of CO$_2$, CO, and H$_2$O.

The first sensor is intended to be used in civil fireplaces. 
Here, the difficulties arise from the simultaneous detection of two different gas species, CO$_2$ and CO, respectively, which are present at largely different concentrations. 
The intended sensor setup and its optical components are described and explained. Furthermore, exemplary measurements of the setup taken at atmospheric conditions are shown.

The second sensor is intended for monitoring EGR rates in IC engines. Here, CO$_2$, being the major combustion product, is the target gas. For each sensor, we describe the field of application and give a brief overview of the setup and its optical components.
Furthermore, a data analysis strategy is presented. An exemplary measurement at a test engine is shown and compared to known properties to validate the data analysis. The article concludes with a summary and outlook.     	
\section{Non-dispersive infrared absorption spectroscopy}
The concentration of the greenhouse gases CO and $\textrm{CO}_2$ in combustion processes can be measured by means of non-dispersive infrared (NDIR) absorption spectroscopy. 
Thereby, infrared radiation is absorbed by the gas molecules as described by the Beer–Lambert–Bouguer law.
The measured radiation intensity is then given by:
\begin{equation}
I(\nu)=I_0(\nu)\mbox{e}^{-\alpha(\nu)\cdot L},
\label{eq:beer_lambert}
\end{equation}
where $I_0$ is the radiation intensity of the source, i.e., without gas in the absorption path, $\alpha$ is the absorption coefficient of the molecules, $\nu$ the light frequency, and $L$ the absorption path length. 
Integration over a frequency interval leads to:
\begin{equation}
\tau=\frac{I}{I_0}= \int^{\nu_{\textrm{max}}}_{\nu_\textrm{{min}}} e^{-\sigma(\nu,p,T)\cdot\rho(p,T)\cdot L}\mbox{d}\nu.
\label{eq:int_lambert}
\end{equation}
Here, $\sigma(\nu,p,T)$ is the frequency, pressure, and temperature dependent absorption cross section and $\rho$ the density of the specific gas.
To describe real absorption measurements, Eq.\,(\ref{eq:int_lambert}) needs to accommodate the systems' transfer function, i.e., the systems' transmission:
\begin{equation}
\tau_{sys}(\nu)=\tau_{filter}(\nu)S_{detector}(\nu)I_{LS}(\nu),
\label{eq:sys}
\end{equation}
with $\tau_{filter}(\nu)$ the transmission spectrum of the filter, $S_{detector}$ the sensitivity of the detector, and $I_{LS}(\nu)$ the spectral intensity of the light source. These are the most common optical components, and this equation can be further expanded to include other used components such as optical fibers.
Combining Eqns. \ref{eq:int_lambert} and \ref{eq:sys}, the normalized transmission $\tau$ in NDIR is then given by:
\begin{equation}\tau=\frac{\int e^{-\sigma(\nu,p,T)\cdot\rho(p,T)\cdot L}\cdot\tau_{sys}(\nu)\mbox{d}\nu}{\int \tau_{sys}(\nu)\mbox{d}\nu}.
\label{eq:norm_ndir}
\end{equation}
\section{Absorption spectrum}
\begin{figure}
	\includegraphics[width=8.3cm]{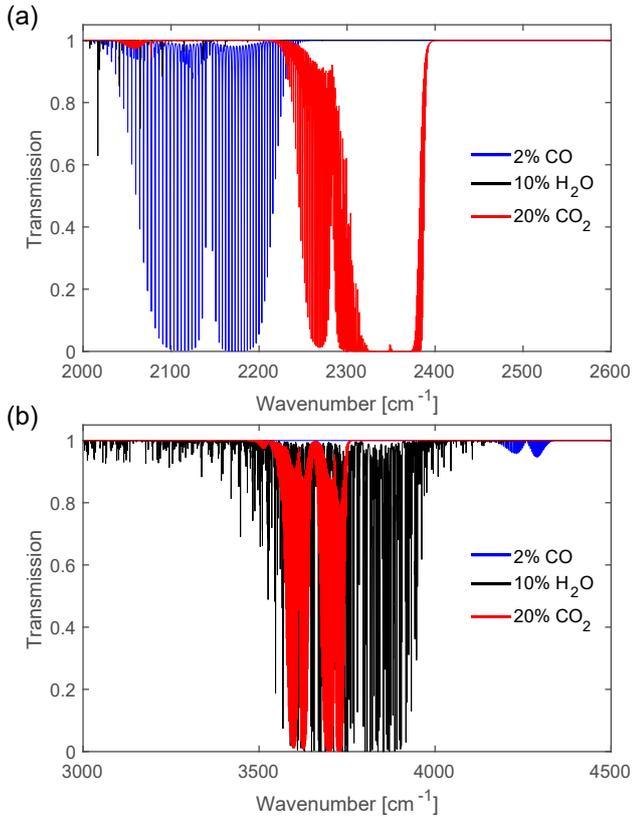}
    \caption{\textbf{(a)} Spectra of exhaust gases common in civilian fireplaces in the range 2000-2500\,cm$^{-1}$ at $T=300$\,K and $p=1$\,bar with an absorption length $L=6$\,cm. Both CO and CO$_2$ have strong absorption bands in this part of the infrared spectral region. \textbf{(b)} The spectral response of CO, CO$_2$, and H$_2$O in the near-infrared (NIR) from 3000-4500\,cm$^{-1}$. The absorption of CO is too low to be used in NDIR spectroscopy, but CO$_2$ and H$_2$O both have strong but overlapping absorption bands.}
    \label{fig:spectra}
\end{figure}

The absorption spectra of the desired gases and their interference with other specimen found in combustion processes are first calculated using the HITRAN database \citep{GORDON20173} and Eq.\,(\ref{eq:int_lambert}).
The results set limits to the wavelength as well as detection range for each gas species and for each setup. 
The absorption spectra for the most relevant constituents are shown in Fig.\,\ref{fig:spectra}.
The concentrations of the carbon oxides used for the calculations are the upper limits allowed in civil fireplaces \citep{VDI-Abgas}. 
H$_2$O concentration is chosen large enough to see any overlaps; concentrations of water vapor in combustion processes can vary extremely with operation conditions.
For the first sensor, intended to be used in civilian fireplaces, the spectral range between 2000 and 2500\,cm$^{-1}$ is chosen, which lies in the mid-infrared (MIR) spectral region, as shown in Fig.\,\ref{fig:spectra}(a). 
\noindent
Here, the rotational-vibrational absorption is very strong, i.e., the absorption coefficients for both CO and CO$_2$ are very large.
The dominant excited vibration for CO is the stretch vibration and for CO$_2$ the $\nu_3$ antisymmetric stretch vibration \citep{gerakines1994infrared}.
Furthermore, there is little to no interference with other gases. 
Figure~\ref{fig:spectra}(b) shows the high-frequency spectral region (NIR) of the absorption spectrum, which is the spectral region chosen for the second sensor.
\noindent
As is shown, there is a strong overlap between CO$_2$ and H$_2$O, which has to be accounted for in the data analysis. 

\section{Sensor Setups}
\subsection{Sensor for use in civilian fireplaces}
This sensor is a combined sensor able to simultaneously detect CO and CO$_2$ and intended to be used by chimney sweepers. 
Therefore, it has to be durable, cost-effective, low-maintenance, easy-to-use, compact, and uphold measurement regularities. These regularities demand the detection of the two carbon oxides in different detection ranges; whereas CO needs to be detected in the range up to 2 vol$\%$ and CO$_2$ up to 20\,vol$\%$.

\begin{figure}
 	\includegraphics[width=8.3cm]{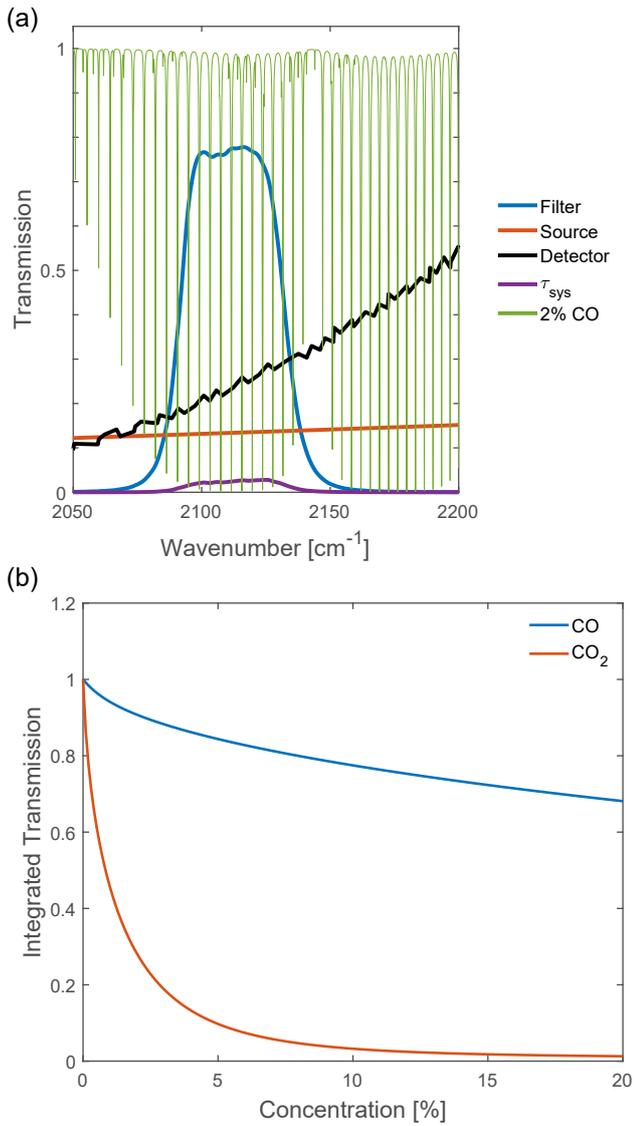}
\caption{\textbf{(a)} Spectral influences on the measured signal. The area beneath $\tau_{sys}$ is the integrated transmission which is equivalent to the measurement signal. \textbf{(b)} Calculated integrated transmission for CO and CO$_2$ for an absorption length of 5\,cm. The difference in the curves shows the need of two different absorption path lengths in the setup to acquire a desirable resolution.}
\label{fig:eff-spek}
\end{figure}

In our setup, we first determined the absorption length for both carbon oxides in order to design a compact sensor for both gas components. 
For this purpose, we simulate the response using the HITRAN data base computing the integrated transmission in Eq.~(\ref{eq:norm_ndir}). 
The transfer function of the sensor ($\tau_{sys}$) is illustrated in Fig.~\ref{fig:eff-spek} (a), where the area beneath $\tau_{sys}$ is the integrated transmission signal. In fact, the recorded signal is very low, which is here a direct result of the emission characteristics of the light source (black body radiation at $2000\, \textrm{K}$) and the sensitivity of the detector in this spectral region.
\noindent
Exemplary, integrated transmissions for different concentrations of CO and CO$_2$ are calculated for an absorption length of $5\,\textrm{cm}$ and shown in Fig.~\ref{fig:eff-spek} (b).
\noindent
From the derived curves one can determine that CO has a much weaker absorption coefficient in comparison to CO$_2$ and thus a higher absorption length (factor of 10) is required for a sufficient sensitivity. 
Furthermore, a large dynamic range of the CO$_2$ sensor extends only up to concentrations of about $5\,\textrm{vol}\%$, being equivalent to a small absorption length for $\mbox{CO}_2$.

A design for a combined sensor for both carbon oxides has to take the aforementioned differences into account. 
The main difficulty lies thereby in the combination of the different measurement ranges where the sensor should show high dynamic response.
For typical sensor applications in civil environment CO$_2$ concentrations range in the $\textrm{vol\%}$ (0.2-20\,vol$\%$), while CO concentrations at the same time lie in the ppm-regime (0-200\,ppm), as the latter is highly toxic and 30\,ppm is the suggested upper limit for working conditions by the BAuA \citep{BAuA2018}.

\begin{figure}
	\includegraphics[width=8.3cm]{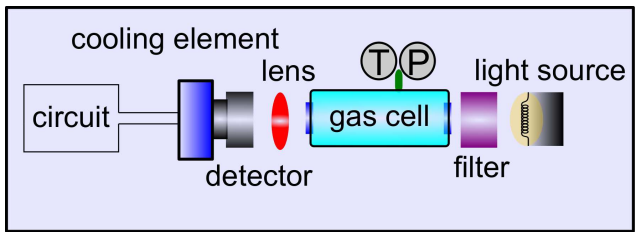}
    \caption{Schematic setup for detection of individual carbon oxides in civilian fireplaces.}
    \label{fig:coco-setup}
\end{figure}

\begin{figure}
 \includegraphics[width=8.3cm]{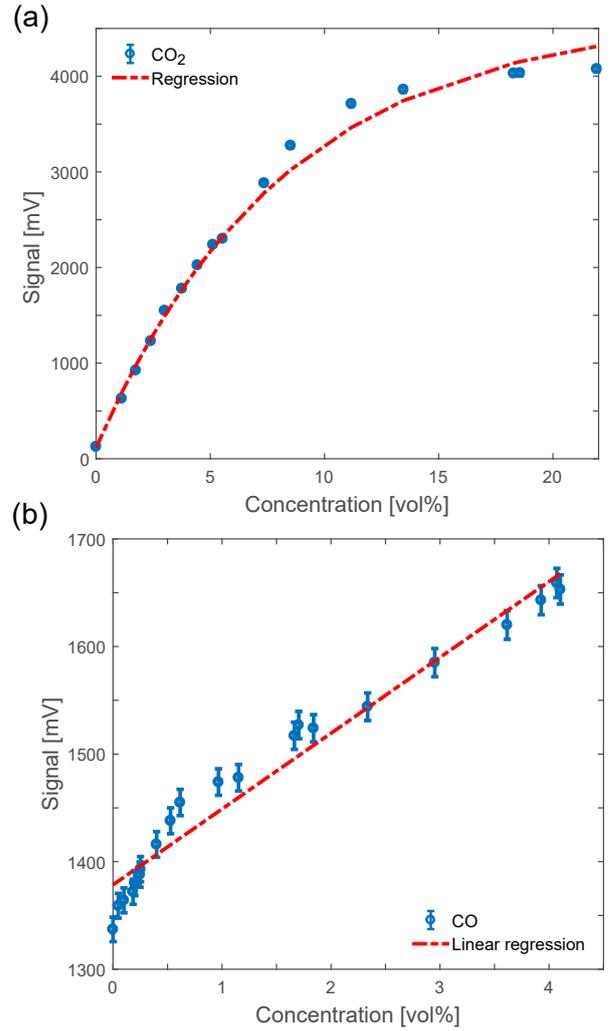}
 	\caption{Measurements of \textbf{(a)} CO$_2$ and \textbf{(b)} CO at atmospheric conditions (1\,bar, 300\,K) and $5\,\textrm{cm}$ absorption path. Included are the regression functions as dotted lines. The measurement is such that decreasing intensity gives a positive signal change.}
    \label{fig:meas-coco}
\end{figure}

A sketch of a setup for each gas is depicted in Fig.\,\ref{fig:coco-setup}.
\noindent
A larger absorption length for CO may be realized by means of a multi-pass cavity, so the sensor dimensions still remain compact.
As light source we chose a bulb with a tungsten filament emitting black-body radiation at $2000\,\textrm{K}$. 
For use as a detector there are two reasonable options: a photo-resistor and a pyroelectric detector. 
The advantage of the latter is its low price, while the former exhibits the better signal-to-noise ratio. Here, a PbSe photo-resistor has been used. 
The circuit to achieve a measurable signal is a Wheatstone bridge with an amplifier circuit.
Exemplary, measurements for both gases are shown in Fig.\,\ref{fig:meas-coco}. 
\noindent
In panel (a), the CO$_2$ sensor shows a saturation behavior. This is due to the saturation of the CO$_2$ transmission filter for its central wavenumbers, as shown in Fig.\,\ref{fig:spectra}(a), so at higher concentrations only the shoulders of the filter spectrum contribute to the signal change.
The regression has the form:
\begin{equation}
f(x)=c-b\cdot e^{-m\cdot x}.
\end{equation}
The sensitivity $s$ of the sensor can be calculated from the derivative of the regression function:
\begin{equation}
s=\frac{\partial f(x)}{\partial x}=mb\cdot e^{-m\cdot x}.
\end{equation}
To give an overview of the sensor attributes the weighted average of the sensitivity is calculated to $\overline{s}=269.75\pm2.88\,\mbox{mV}\,\mbox{vol}\%^{-1}$. 
With a digital resolution $R_{\textrm{digit}}$ of 1\,mV the resolution $\Delta r$ can be calculated as
\begin{equation}
\Delta r=\frac{R_{\textrm{digit}}}{s},
\end{equation}
which leads to an overall CO$_2$ resolution of $\Delta r=60\pm1$\,ppm. Hazardous workplace environment is defined above 0.5\,vol$\%$ \citep{BAuA2018}, so the resolution is usable for civil fireplaces.

In Fig.\,\ref{fig:meas-coco}(b), the CO sensor shows an approximately linear dependence as expected from Fig.\,\ref{fig:eff-spek} (b). 
The fluctuations visible in the measurement are a result of the cooling routine of the sensor. 
The sensitivity is $\overline{s}=7.14\pm0.04\,\mbox{\textmu V}\,\mbox{ppm}^{-1}$, which results in a resolution of $\Delta r=139.90\pm6.58$\,ppm. 
As already mentioned, a CO concentration over 30\,ppm is hazardous to be exposed to for a longer time. Therefore, the sensitivity of CO needs to be enhanced to achieve a higher resolution. 
This can be accomplished on the one hand by the use of better detectors, which would result in more expensive sensors and on the other hand, a larger absorption path length. 
The development of the latter, i.e., a multi-pass configuration for the CO absorption measurements is a task of current research.  

\subsection{Exhaust gas sensor for IC engines}
While the sensor described above represents a cost-sensitive down market application, NDIR spectroscopy might also be used in more demanding environments in research and development. In the following section, we present a second sensor capable of quantification of residual gas concentrations in internal combustion (IC) engines. In contrast to the fireplace exhaust sensor described above, an IC engine requires measurements under highly dynamic conditions with pressures and temperatures ranging from $1-40\,\textrm{bar}$ and $300-1000\,\textrm{K}$, respectively. Moreover, the sensor needs to offer a high temporal resolution (at least <1\,ms). Since the field of application is less cost-sensitive, the requirements can be met by the use of high-end components. Moreover, an accurate quantification over the wide range of conditions requires a more sophisticated data analysis that uses the well-known spectroscopic properties of the molecules. 

\begin{figure}
\includegraphics[width=8.3cm]{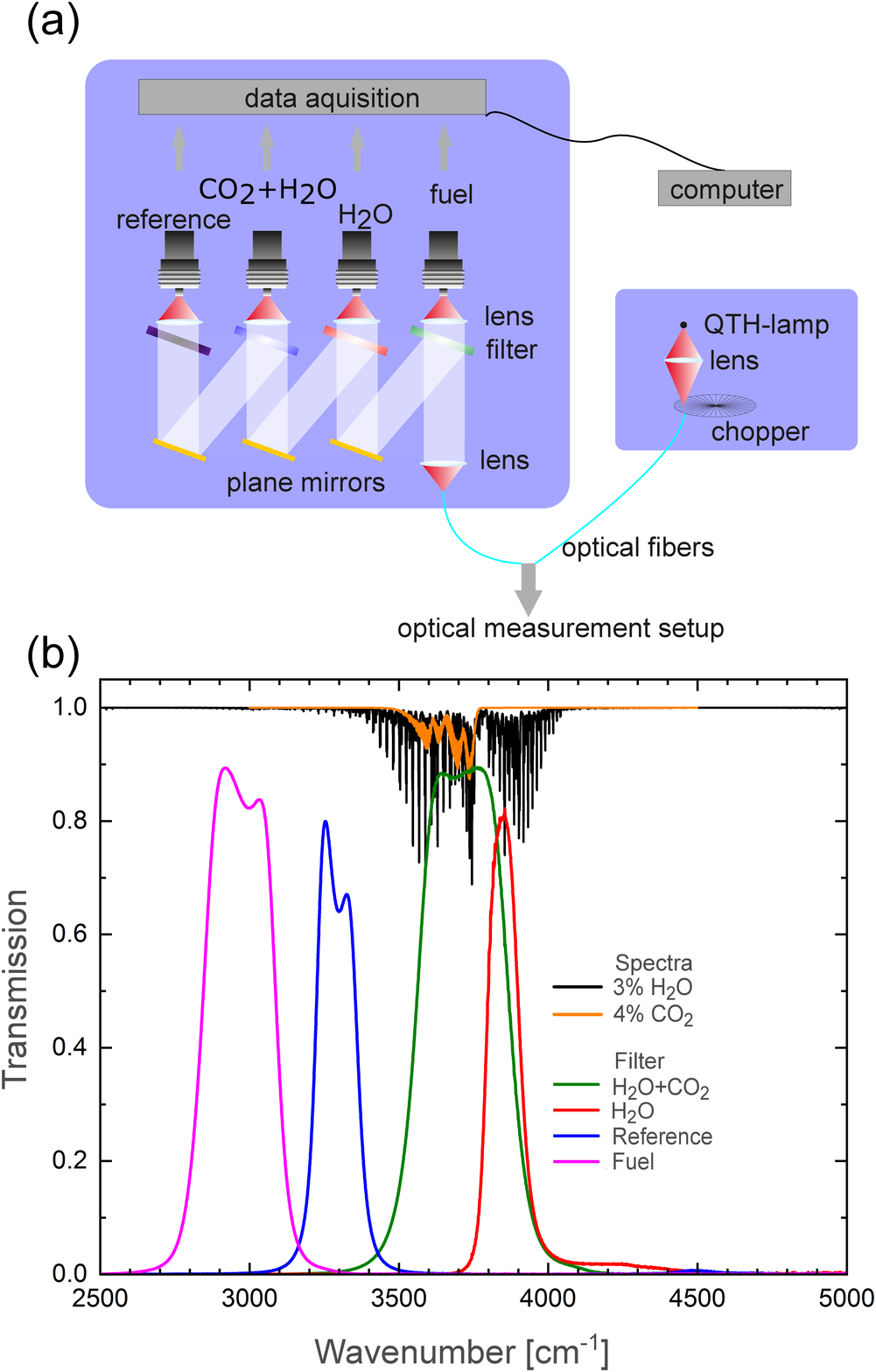}
\caption{\textbf{(a)} Schematic of the ICOS sensor system consisting of different measurement channels. \textbf{(b)} Transmission spectra of the bandpass filters used in the ICOS sensor system as well as the spectra for H$_2$O and CO$_2$. Note that the serial assembly causes the bandpass filter influencing each other.}
\label{fig:omega-e}
\end{figure}

The sensor system used in this work is a modification of the Internal Combustion Optical Sensor (ICOS) for LaVision GmbH and has been described extensively elsewhere (\cite{Grosch_2014}; \cite{Golibrzuch2017}; \cite{Bauke2018}). The schematic layout of the system is shown in Fig.\,\ref{fig:omega-e}(a).
Briefly, the system consists of a broad-band light source (150\,W quartz-tungsten-halide (QTH) lamp), a spark plug sensor probe and a detection unit. The light from the QTH lamp is modulated by a 30\,kHz chopper wheel, which determines the maximum time resolution to about $33\,\textrm{µs}$, and coupled into a ZrF$_4$ mid-infrared fiber. The time resolution is sufficient to enable crank-angle resolution of single cycles in IC engines up to 5000 rpm.
The ZrF$_4$ fibers guide the light to the spark plug probe and back to the detection unit. Inside the probe, sapphire fibers guide the light to the detection volume where it is reflected back by a concave mirror located in a stainless steel cage holder. The absorption path is 0.96 cm. The sapphire fibers are necessary to withstand the high temperature during fired engine operation, but limit detection to wavelengths $\lambda\,<\,3.6\,\textrm{µm}$. 
Inside the detection unit the light passes a cascaded array of Mercury-Cadmium-Telluride (MCT) detectors equipped with different bandpass filters. An overview of bandpass filters used in this work is shown in Fig.\,\ref{fig:omega-e}(b). The system consists of a filter for fuel concentration measurements, utilizing the absorption of C-H stretch vibrations of hydrogen carbons around 3100\,cm$^{-1}$ as well as two detection channels for water and CO$_2$ and an “offline” reference filter situated in a spectral range with negligible absorption of any present molecular species. The reference filter serves as a correction for signal disturbances due to, e. g., beam steering or particles in the beam path. Note that the effective transmittance curves differ from the raw ones due to the serial assembly of the filters.
Fuel concentration measurements are beyond the scope of this work but have been demonstrated for gasoline (\cite{Grosch_2007}; \cite{Grosch_2010}; \cite{Grosch_2011}) as well as methane-fueled engines ( \cite{Bauke2017}; \cite{Golibrzuch2017}; \cite{Bauke2018}; \cite{Kranz2018}) and we will focus on the quantification of residual gas, i.e. CO$_2$ and H$_2$O.

Since the extreme conditions in an IC engine require the use of sapphire fiber to guide the light into the combustion chamber, CO$_2$ detection at 2400\,cm$^{-1}$ is impossible. Therefore, the detection is limited to the weaker absorption band around 3700\,cm$^{-1}$ which, however, is completely blended by water absorption. Consequently, a strategy is required to correct the influence of water vapor. As visible from the spectra shown in Fig.\,\ref{fig:omega-e}(b), the H$_2$O absorption covers a much larger spectral range than CO$_2$. Therefore, a second filter that is only sensitive to water is used to determine the H$_2$O amount independently. Additional complexity arises, since dynamic changes in pressure and especially temperature need to be accounted for.

\subsubsection{Data analysis}
In order to enable quantification in the large range of experimental conditions, the data analysis procedure relies on the complete description of the molecule's spectroscopic properties, making use of the HITRAN database \citep{GORDON20173}, as well as the spectral influences of the optical system (see Eq.\,(\ref{eq:norm_ndir})).
This procedure has been recently described in more detail in (\cite{Golibrzuch2017}; \cite{Bauke2018}) for methane and might be applied for CO$_2$ and H$_2$O accordingly.

Briefly, we use the spectroscopic constants from HITRAN to calculate the transmission of the H$_2$O+CO$_2$ and the H$_2$O detection channels for different CO$_2$ and H$_2$O concentrations as a function of temperature and pressure ranging from $250-1000\,\textrm{K}$ and $0.1-40\,\textrm{bar}$, respectively.
The computed data are then used to build a 3-dimensional look-up-table that links the measured transmission to the corresponding molecules’ density for different pressures and temperatures.

While, temperature and pressure information for the civil fireplace sensor is easily accessible, the IC sensor faces significant problems, especially regarding the temperature in the measurement volume.
Time-resolved pressure measurements are usually available at engine test stations, but temperature is usually unknown due to the comparable low speed of standard probes. Moreover, the temperature in the measurement volume, surrounded by a metal cage, can differ strongly from the temperature commonly calculated by thermodynamic models (\cite{Golibrzuch2017}; \cite{Kranz2018}).
Therefore, temperatures need to be estimated from modified thermodynamic models or measured by other spectroscopic techniques (\cite{Luong_2008}; \cite{Werblinski2017}); the latter might however require the use of an additional probe.
The possibility to determine temperatures using NDIR probing different spectral regions of an absorption band simultaneously has thereby been demonstrated recently for the case of methane (\cite{Golibrzuch2017}; \cite{Bauke2018}). In case of CO$_2$/H$_2$O, this would however require the use of third filter, which would further raise the complexity of the system and the data analysis procedure. 
Given that all information for quantification (transmission, pressure, and temperature) are available, the remaining challenge is to disentangle absorption due to $\textrm{CO}_2$ and $\textrm{H}_2\textrm{O}$, respectively. 

The sensor system offers transmission information in a spectral region with only $\textrm{H}_2\textrm{O}$ absorption
lines as well as for a region with combined  $\textrm{CO}_2$ and $\textrm{H}_2\textrm{O}$ absorption. 
In a first approximation, we assume that the absorption in the overlapping region can be described as the product of transmission caused by H$_2$O and CO$_2$:
\begin{equation}
\tau_{\tiny{\mbox{H}_2\mbox{O}+\mbox{CO}_2}}=\tau_{\tiny{\mbox{CO}_2}}(T,p,\rho_{\tiny{\mbox{CO}_2}}) \times \tau_{\tiny{\mbox{H}_2\mbox{O}}}(T,p,\rho_{\tiny\mbox{H}_2\mbox{O}}).
\label{eq:tauCO2_H2O}
\end{equation}
It is important to note here, that $\tau_{\tiny{\mbox{H}_2\mbox{O}+\mbox{CO}_2}}$, $\tau_{\tiny{\mbox{CO}_2}}$ and $\tau_{\tiny{\mbox{H}_2\mbox{O}}}$ are broadband transmittance values. While, Eq.\,(\ref{eq:tauCO2_H2O}) would be completely valid for frequency dependent transmittance, it is only an approximation for direct multiplication of integrated broadband values. 

\begin{figure}
		\includegraphics[width=8.3cm]{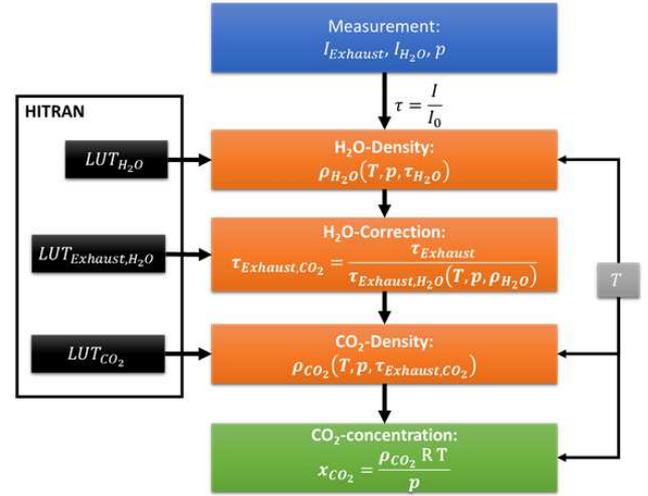}
    \caption{Schematic overview of the data analysis procedure for determination of CO$_2$ concentrations in IC engines by NDIR.}
    \label{fig:schematic}
\end{figure}

Since the water density can be determined from the pure H$_2$O signal in the second detection channel,
$\tau_{\tiny{\mbox{H}_2\mbox{O}}}$ can be calculated and the CO$_2$ density remains the only unknown variable to be determined. Figure\,\ref{fig:schematic} shows a schematic overview of the data analysis procedure applied in this work. Note that the temperature information effects different points of the procedure.

\begin{figure}
\includegraphics[width=8.3cm]{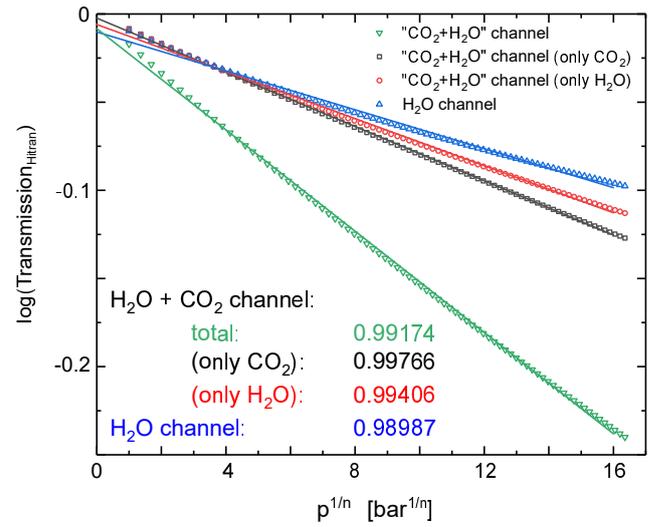}
\caption{HITRAN simulation of the logarithmic transmission as a function of $p^{\frac{1}{n}}$ for H$_2$O and CO$_2$ absorption in the respective detection channels for a typical polytropic compression in an IC engine ($T_0=300K$, $p_0=1 bar$, $n=1.32$).}
\label{fig:I0_extrapol}
\end{figure}

Another important issue, is the determination of $I_0$, i.e. the signal without absorption species in the beam path. In contrast to fuel concentration measurements, where $I_0$ can be determined before fuel enters the combustion chamber (\cite{Golibrzuch2017}; \cite{Kranz2018}), water and CO$_2$ are always present in ambient air as well as exhaust gas. Therefore, the detector signal received at the lowest gas density always contains some absorption. In order to eliminate this effect, we developed a method for $I_0$ determination by extrapolation to $p=0$. In an IC engine, temperature and pressure usually follow an polytropic compression:
\begin{equation}
T=\frac{p}{p_0} \cdot {(\frac{T}{T_0}})^{\frac{n-1}n}
\label{eq:poly_compr}
\end{equation}
where $p_0$ and $T_0$ are the pressure and temperature prior to compression and $n$ is the polytropic coefficient. Assuming that Beer–Lambert–Bouguer law (Eq. (\ref{eq:int_lambert})) is also approximately valid for integrated transmission using an 'integrated absorption cross-section', $\sigma$:
\begin{equation}
I_{int} \approx I_{0,int} \mbox{e}^{-\sigma(p,T) \cdot \rho(p,T) \cdot L}
\label{eq:beer_lambert_int}
\end{equation}
Eq. (\ref{eq:beer_lambert_int}) can be linearized to:
\begin{equation}
ln(I) \approx ln(I_{0}) -\sigma(p,T) \cdot \rho(p,T) \cdot L
\label{eq:log_beer_lambert_int}
\end{equation}
For a polytropic compression the gas density can be expressed using ideal gas law as:
\begin{equation}
\rho(p,T)=\frac{p}{R \cdot T}
\end{equation}
and with Eq. (\ref{eq:poly_compr}):
\begin{equation}
\rho(p,T) = \frac{{p_0}^{\frac{n-1}n}}{R \cdot T_0} \cdot p^{1-\frac{n-1}n}=const.\cdot p^{\frac{1}n}
\label{eq:poly_rho}
\end{equation}
Consequently, $I_0$ can be extracted as the intercept of a linear fit to $ln(I)$ as a function of $p^{\frac{1}{n}}$. An accurate $I_0$ determination requires that the approximations of Eq. (\ref{eq:log_beer_lambert_int}) and the integrated absorption cross-section $\sigma$ being independent of pressure and temperature are valid. Fig. \ref{fig:I0_extrapol} shows  HITRAN simulation of the signals in the H$_2$O+CO$_2$ and the H$_2$O detection signals for a typical polytropic compression. The data shows that the approximations made above hold reasonably well for most signals and can be judged from the intercept of the linear fits being close to 0 ($I_0=1$). The best results are expected if only CO$_2$ is present since the respective bandpass filter covers the complete CO$_2$ absorption band. In case of H$_2$O, the sensor sees only a part of the absorption and temperature effects due to redistribution of rotational states are more relevant.

\subsubsection{Quantification of CO$_2$ and H$_2$O in an IC engine}
\begin{figure}
	\begin{center}
		\includegraphics[width=8.3cm]{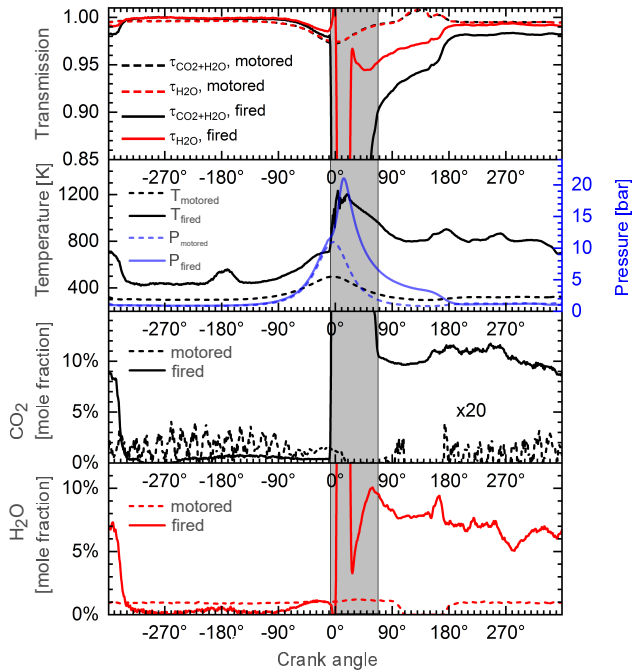}
	\end{center}
	\caption{Measurements of CO$_2$ and H$_2$O concentrations in a methane-fueled IC engine. Dashed lines: motored operation with laboratory air. Solid lines: fired operation with methane port-fuel-injection at a global $\lambda=1$ (stoichiometric combustion). The gray area indicates time of combustion which is excluded from the data interpretation due to partial saturation of the detectors from the flame emissions.}
    \label{fig:engine-icos}
\end{figure}

We test the system and data analysis procedure in a methane-fueled IC engine under motored (pure air) and fired operation under stoichiometric conditions. Details of the engine used in this work are given in \cite{Kranz2018}.
Figure\,\ref{fig:engine-icos} gives an overview of the respective results averaged over 100 engine cycles. Panel (a) shows the measured transmission signals under motored (dashed lines) and fired (solid lines) conditions for both channels, CO$_2$+H$_2$O (black) and H$_2$O (red), respectively. 
Panel (b) shows corresponding temperature (black) and pressure (blue) data. 
Note that the temperature data were obtained in a separate measurement using spectrally resolved water absorption measurement from an ICOS-Temperature system (LaVision GmbH) (\cite{Bauke2018}; see \cite{Werblinski2016}; \cite{Werblinski2017} for operation principle). 
Panel (c) and (d) show the results for CO$_2$ and H$_2$O concentrations obtained using the data analysis described above. Under motored engine operation, we determine a CO$_2$ concentration of about 0.05\,$\%$, which is in good agreement with ambient CO$_2$ concentration of 0.04\,$\%$, but close the detection limit of the system. The water concentration is determined to approx. 1\,$\%$ corresponding to about 40\,$\%$ humidity at 293\,K \citep{Wexler1976}. 
Consequently, the motored data indicates that the analysis algorithm and modeling yields results over a wide range of pressure and temperature, which are consistent with typical ambient conditions. However, more detailed validation experiments in e.g. static pressure cells or a rapid compression engine are needed in order to evaluate achievable accuracy and precision.
Nevertheless, the promising motored results enable a first evaluation of the system and fired engine operation conditions (solid lines in Fig.\,\ref{fig:engine-icos}). The IC engine was operated with methane port-fuel-injection under global stoichiometric conditions. At the beginning of the engine cycle at $-360^{\circ}\,\textrm{CA}$, we obtain CO$_2$ and H$_2$O concentrations of about 9\,$\%$ and 7\,$\%$, respectively. After opening of the intake valves at -334°CA, a mixture of ambient air and methane enters the combustion chamber, leading to a strong decrease in the exhaust gas concentrations. During compression from $-180^{\circ}\,\textrm{CA}$ to $-10^{\circ}\,\textrm{CA}$, air, fuel, and residual exhaust gas undergo a process of mixture formation resulting in final CO$_2$ and H$_2$O concentrations of 0.4\,$\%$ and 1\,$\%$, respectively.  
After ignition at $-10^{\circ}\,\textrm{CA}$, the transmission signals exhibit a steep decrease due to flame emission partially saturating the detectors (gray area). The region is therefore excluded from the data analysis. 
After combustion, the temperature remains at about 800\,K with CO$_2$ and H$_2$O concentrations of about 10\,$\%$ and 8\,$\%$. 
From this data, we can estimate the EGR rate to be approximately $5\,\%$. The EGR rate might also be estimated from pressure, temperature and volume using ideal gas law. 
Before opening of the intake valve at -334°CA ($p_{\textrm{IVO}}=0.96\, \textrm{bar}$, $T_{\textrm{IVO}}=600\,\textrm{K}$, $V_{\textrm{IVO}}= 66.1\,\textrm{ccm}$), a remaining gas amount of $n_{\textrm{IVO}}=1.3 \times 10^{-3}\, \textrm{mole}$ can be estimated. 
After intake valve closing at -184°CA ($p_{\textrm{IVC}}=0.84\,\textrm{bar}$, $T_{\textrm{IVC}}=335\textrm{K}$, $V_{\textrm{IVC}}=594\,\textrm{ccm}$), the total gas amount raised to $n_{\textrm{IVC}}=1.8 \times 10^{-2}\,\textrm{mole}$. 
Comparison $n_{\textrm{IVO}}$ to $n_{\textrm{IVC}}$ gives an EGR rate of $7\,\%$, consistent with the sensor data.

\section{Summary and outlook}  
We presented here two gas sensors based on non-dispersive infrared spectroscopy for high and low tech. 
We outlined the development of these sensors, one intended for civil fireplaces with the ability to detect CO as well as CO$_2$ and the other for IC engines, capable of CO$_2$ and water vapor detection.
Potential spectral regions for the detection of CO$_2$ were identified between 3400 and 4000\,cm$^{-1}$ and 2200 to 2400\,cm$^{-1}$, whereas the former strongly overlaps with water absorption bands. 
This disadvantage leads to the necessity of a water absorption channel and additional calculations to separate CO$_2$ and H$_2$O. 

Additionally, at the lower frequency region from 2200 to 2400\,cm$^{-1}$ CO has absorption bands, which only have a negligible overlap with the CO$_2$ absorption bands. This enables the simultaneous determination of the concentrations in a single sensor, which is suitable for civil fireplaces based on non-dispersive infrared spectroscopy. 
The absorption behavior of CO and CO$_2$ are compared and their optimal absorption lengths were discussed, whereas CO needs a long and CO$_2$ needs a short absorption length. Furthermore, we discussed possible optical components for the sensor. The final main components are a thermal broadband emitter, optical filter, and a PbSe photo resistor.
The presented sensor is capable of a resolution of 60\,ppm for CO$_2$ and 140\,ppm for CO at an absorption length of $L=5\,\textrm{cm}$.
Due to the usage of a PbSe detector, temperature has a tremendous influence on the sensor, which can especially be observed during the CO measurement. 
Here, the cooling routine of the sensor is visible in oscillations of the measurement points in time. 
To minimize the influence of temperature and get rid of thermal oscillations, an improved cooling routine has to be implemented. Furthermore, it is advisable to use another material as a detector, as lead (Pb) may be further regulated by the European Union. 
Here, InAsSb or pyroelectric sensors may be utilized. It is also possible to use a different spectral filter for the detection of CO$_2$, as for higher concentrations oversaturation is encountered. It can be advisable to use a CO$_2$ sensor on the flanks of the CO$_2$ absorption, so a linear measurement might be possible.

The spectral region between 3000 and 4500\,cm$^{-1}$ and its CO$_2$ absorption is utilized for a sensor intended for IC engines due to limitations for mid-infrared fiber guides. To address the overlap of H$_2$O and CO$_2$ in this spectral region, the detector consists of multiple detection channels built like a cascade to achieve a single detection channel of H$_2$O and a compound channel of the mixture of H$_2$O and CO$_2$. A calculation routine utilizing look-up tables is presented to achieve a single water and CO$_2$ signal. 
This sensor is capable of high time resolutions up to 33\,µs and faces huge challenges due to a harsh and highly dynamic (temperature and pressure) environment. 
We demonstrated an application of the system to an IC engine under motored and fired operation. The results under motored operation were consistent with typical ambient condition. For fired conditions, these data could be used to calculate the EGR rate which was in agreement with thermodynamic estimations. Nevertheless, further validation   of the sensor and data analysis under more controlled conditions (e.g. in static pressure cells or a rapid compression engine) is required to determine its accuracy and precision over a wide range of temperatures and pressures.


\dataavailability{The experiments and results shown in this publication
	are strongly industry-related research. We explain our experimental
	preparations and analysis steps in great detail in this
	work and are available for questions. Please understand that therefore
	we do not publicly provide the underlying data and Matlab code
	used for analysis. Given individual requests by fellow researchers,
	we will of course consider making parts of the data available.} 






\noappendix       







\authorcontribution{All authors contributed equally to the experiments and to the manuscript preparation.} 

\competinginterests{The authors declare that they have no conflict of interest.} 


\begin{acknowledgements}
The authors gratefully acknowledge financial support through the Federal Ministry of Education and Research (BMBF, Germany) FKZ: 13N13035 and the Federal Ministry for Economic Affairs and Energy (BMWi, Germany) FKZ: ZF4060502WM6. 
\end{acknowledgements}







\bibliographystyle{copernicus}
\bibliography{BIB}

\end{document}